\documentclass[5p,times,twocolumn,authoryear]{elsarticle}
\usepackage[T1]{fontenc}
\usepackage{amssymb}
\PassOptionsToPackage{hyphens}{url}
\usepackage{hyperref}
\usepackage[hyphens]{url}
\usepackage{natbib}
\usepackage{longtable}
\usepackage{tabularx}
\usepackage{multirow}

\usepackage[final]{changes}

\newcommand{\nsobjtyp}{https://ivoa.net/rdf/object-type/\#}
\newcommand{\nstarcls}{https://voparis-ns.obspm.fr/rdf/epn/2.0/target-class/\#}
\newcommand{\nsuat}{http://astrothesaurus.org/uat/}
\newcommand{\nsivoauat}{https://ivoa.net/rdf/uat/\#}
\newcommand{\nsskos}{http://www.w3.org/2004/02/skos/core\#}
\newcommand{\nsivoasem}{https://ivoa.net/rdf/ivoasem/\#}
\newcommand{\nsdct}{http://purl.org/dc/terms/}

\newcommand{\objtyp}[1]{\href{\nsobjtyp#1}{\texttt{objtyp:#1}}}
\newcommand{\tarcls}[1]{\href{\nstarcls#1}{\texttt{tarcls:#1}}}
\newcommand{\uat}[1]{\href{\nsuat#1}{\texttt{uat:#1}}}
\newcommand{\ivoauat}[1]{\href{\nsivoauat#1}{\texttt{ivoauat:#1}}}
\newcommand{\skos}[1]{\href{\nsskos#1}{\texttt{skos:#1}}}
\newcommand{\ivoasem}[1]{\href{\nsivoasem#1}{\texttt{ivoasem:#1}}}
\newcommand{\dct}[1]{\href{\nsdct#1}{\texttt{dct:#1}}}

\journal{Astronomy $\&$ Computing}

\newcommand{\hdoi}[1]{\href{https://doi.org/#1}{doi:#1}}

\begin{document}

\begin{frontmatter}
\title{OntoPortal-Astro, a Semantic Artefact Catalogue for 
Astronomy}

\author[1]{Baptiste Cecconi\corref{cor1}}
\ead{baptiste.cecconi@observatoiredeparis.psl.eu}
\cortext[cor1]{Corresponding author}
\author[1]{Laura Debisschop}
\author[2]{S\'ebastien Derri\`ere}
\author[2]{Mireille Louys}
\author[4]{Carmen Corre}
\author[5]{Nina Grau}
\author[6,7]{Cl\'ement Jonquet}

\affiliation[1]{organization={{LIRA, Observatoire de Paris, Universit\'e PSL, 
CNRS, Sorbonne Universit\'e, Universit\'e Paris Cit\'e}}, addressline={5 Place 
Janssen}, city={Meudon}, postcode={92190}, country={France}}
\affiliation[2]{organization={Observatoire Astronomique de Strasbourg, CNRS-UMR7550, Universit\'e de Strasbourg}, addressline={11, rue de l'Universit\'e}, city={Strasbourg}, postcode={67000}, country={France}}
\affiliation[4]{organization={INRAE, DipSO}, city={49070 Beaucouz\'e},  country={France}}
\affiliation[5]{organization={CODATA, the Committee on Data of the International Science Council}, city={Paris}, country={France}}
\affiliation[6]{organization={MISTEA, INRAE, Institut Agro de Montpelier}, city={Montpellier}, country={France}}
\affiliation[7]{organization={LIRMM, CNRS, University of Montpellier}, 
city={Montpellier}, country={France}}

\begin{abstract}
The astronomy communities are widely recognised as mature communities for their 
open science practices. However, while their data ecosystems are rather advanced 
and permit efficient data interoperability, there are still gaps between these 
ecosystems. Semantic artefacts (SAs) ---e.g., ontologies, thesauri, vocabularies 
or metadata schemas--- are a means to bridge that gap as they allow to 
semantically described the data and map the underlying concepts. The increasing 
use of SAs in astronomy presents challenges in description, selection, evaluation, 
trust, and mappings. The landscape remains fragmented, with SAs scattered across 
various registries in diverse formats and structures --- not yet fully developed 
or encoded with rich semantic web standards like OWL or SKOS --- and often with 
overlapping scopes. Enhancing data semantic interoperability requires common 
platforms to catalog, align, and facilitate the sharing of FAIR 
(Findable, Accessible, Interoperable and Reusable) SAs. In the frame of the 
FAIR-IMPACT project, we prototyped a SA catalogue for astronomy, heliophysics 
and planetary sciences. This
exercise resulted in improved vocabulary and ontology management in the 
communities, and is now paving the way for better interdisciplinary data 
discovery and reuse. This article presents current practices in our discipline, 
reviews candidate SAs for such a catalogue, presents driving use cases and the 
perspective of a real production service for the astronomy community based on 
the OntoPortal technology, that will be called OntoPortal-Astro. \end{abstract}


\end{frontmatter}
\section{Introduction}
The set of resources in the so-called ``virtual observatory'' 
\citep{2000A&AS..143....1G,2009EM&P..104..315H,
2015A&C....11..190H,2018P&SS..150...65E,2020DatSJ..19...22E,mampaey:2025} 
forms a network of related data products, 
catalogues, services and tools allowing researchers to discover, access
and process data for their research. One example is the International Virtual 
Observatory Alliance\footnote{\url{https://ivoa.net}} (IVOA) framework 
\citep{arviset2018vopowerfultoolglobal}, which proposes a registry 
\citep{2019ivoa.spec.1011D} of resources gathering endpoints for 
services, datasets and catalogues, and is used by tools to discover and provide 
access to resources. This astronomy framework is an example of an open 
science ecosystem. 

\subsection{Astronomy communities}
The astronomy community is mainly composed of three semantics sub-communities: 
celestial astronomy\footnote{In this paper, we use the term ``celestial 
astronomy'' to refer to any topics outside the Solar System. Note that it's not
a commonly accepted definition for this concept, but we use it for the sake of 
simplicity.} (objects are referenced to with their fixed sky coordinates, e.g., 
stars, galaxies, cosmology, etc); heliophysics (the study of the Sun, 
and the plasma environments throughout the Solar System); planetary sciences 
(the study of the Solar System objects, e.g., planets, comets, asteroids, etc). 
Each of these sub-communities have developed interoperability frameworks and
semantic ecosystems, which are rather siloed up to now, compromising the 
semantic and technical interoperability of those ecosystems.

The main astronomy sub-community is the celestial astronomy community. Its data 
management infrastructure is organised around the IVOA, which is 
interoperability driven. The IVOA was founded in 2002, and is now composed of 23
national groups (at the time of writing), which act as relays between the 
community developments and the international cooperation level. The IVOA gathers 
twice per year with the so-called ``IVOA Interop'' conferences. With this fully 
bottom-up framework, the IVOA defines schemas, protocols and vocabularies, which 
are respectively managed by the Data Models, Data Access Layers, and Semantics 
working groups (WGs). In addition, a registry (managed by the IVOA Registry WG) 
provides a catalogue of IVOA compliant resources. Hence, this enables a stable 
and well-established ecosystem adopted by the vast majority of astronomy 
archives, data centres and tool developers. As a result, scientists are using 
applications, such as Aladin \citep{2000:aladin} or TOPCAT 
\citep{2005:topcat,2011ascl.soft01010T}, which are fully relying on IVOA tooling
without seeing any detail of the underlying interoperability layers.

The second sub-community is the heliophysics community. It is organised around 
the International Heliophysics Data Environment 
Alliance\footnote{\url{https://ihdea.net}} \citep[IHDEA,][]{Masson:2024fc}, 
which is distribution\footnote{In the sense of DCAT (Data Catalog Vocabulary), 
see \url{https://www.w3.org/TR/vocab-dcat-3/\#Class:Distribution}} driven. 
The IHDEA was formed in 2018 by nine institutions (from laboratories 
to space agencies) willing to foster collaboration and standardisation of the 
heliophysics data ecosystem. However, the most of heliophysics community 
standards predates the inception of the IHDEA. The Space Physics Archive 
Search and Extract \citep[SPASE,][]{Roberts:2018bi} is the main heliophysics 
information model, which development started in 2002. It is 
maintained by the SPASE Metadata Working Team. It defines a rather complete 
and evolutive way to describe space physics observational datasets, model 
runs, instruments or repositories. For each dataset, it is possible 
to define the various ways of the accessing the data. It is used by 
several major heliophysics data centres \citep[NASA/SPDF, CNES/CDPP, IUGONET, 
see e.g.,][]{Abe:2024ty,HPDataPolicy1.2,Roberts:2018bi} for their internal data 
model. Other metadata guidelines are available to this community (such as the 
ISTP metadata for data stored in CDF \citep{whipple_international_1995,
cdf-istp-spec}, or the SOLARNET metadata for data stored in FITS 
\citep{std:FITS,solarnet_zenodo}). Data access protocols are also standardised, 
with, e.g., the HAPI protocol \citep{Weigel:2021vf}. The IHDEA ecosystem is 
not as interoperable as the IVOA one. While some tools are providing a very 
simple access to heliophysics datasets (such as the 
AMDA\footnote{\url{https://amda.irap.omp.eu}} \citep{2021:amda} or the 
CDAWeb\footnote{\url{https://cdaweb.gsfc.nasa.gov}} web interfaces), using IHDEA 
services or resources often require a deep understanding of interoperability 
interfaces and data formats. However ongoing developments are aiming at building 
tighter links with IVOA and improve open science in heliophysics.

The third sub-community is the planetary science community. It is mainly 
organised around the International Planetary Data 
Alliance\footnote{\url{https://planetarydata.org}} (IPDA), which 
is archive driven. The IPDA was formed in 2006 and is now composed of 13 
members, which are mostly space agencies. The primary role of the IPDA is to 
develop and maintain the NASA Planetary Data System (PDS) \citep{Hughes:2017bt}, 
which is an advanced information model for planetary exploration science 
archives. The PDS is used by all major space agency archive with planetary 
exploration missions: NASA/PDS, ESA/PSA, JAXA, ISRO, etc. Moreover, planetary 
scientists studying the planetary surfaces are using the OGC\footnote{
\url{https://www.ogc.org/}} (Open Geospatial 
Consortium) ecosystem, developed by the geoscience community (Earth 
observation). In 2021, a small group of planetary scientist formed the 
Planetary Domain Working Group, which role is to adopt the OGC standards for 
the study of planetary surfaces. Technical interoperability 
between OGC and IVOA has been studied for some years 
\citep{HARE201836,Marmo:2018,minin:2018,erard:epsc:2024}. 
However, semantic interoperability still needs to be implemented.

Each astronomy sub-community data framework implements various pieces 
of an open science ecosystem with a different goal: celestial astronomy 
focuses on interoperability; heliophysics, on distribution; and planetary 
sciences, on long term archive. This also implies that different ways of 
registering, connecting, publishing and accessing resources and their 
metadata.

\subsection{Linked-data architecture}

The linked-data architecture has been defined at the inception times of the 
World Wide Web \citep{tbl-semantic-web}. The original 
framework --- called RDF: Resource Description Framework --- 
\citep{note:rdfprimer} developed and published by the W3C proposes a description 
of resources and their relations, based on triples: a subject, a predicate and an 
object. The triples can be seen as sentences, the predicate being the verb. The 
set of triples creates an oriented and tagged graph of nodes. Nodes are resources 
(referred to as URI: Unified Resource Identifier). Predicates (or properties) 
designate the links connecting the ressources. The classes, their 
instances and the properties are usually defined in standard vocabularies or 
ontologies capturing a certain level of semantics that a machine will then rely 
to query and reason over the data. 

In the present case, this graph of nodes, metadata and relations is 
capturing knowledge about scientific resources, hence we call 
it a ``scientific knowledge graph''. Keeping this graph of resources usable and
interoperable while it grows, requires 
semantic and technical coordination and synchronisation, so that terms 
are globally understood by the tools and actors providing the content and 
enabling the access. This semantic and technical coordination is achieved by 
methods, technologies and tools coming from a domain called, the Semantic Web 
for which semantics artefacts, are at the core \citep{semantic-artefact}. A 
semantic artefact\footnote{In \citet{semantic-artefact}, a ``semantic 
artefact'' is defined as \emph{a machine-actionable formalisation 
(represented using appropriate formats and serialisations, including RDF and 
non-RDF standards) of a conceptualisation, enabling sharing and reuse by 
humans and machines}. In this paper, the term is used to refer to both 
machine-actionable and non-machine-actionable resources.}
(hereafter referred to as SA) --- a broader term to include 
ontologies, terminologies, taxonomies, thesauri, vocabularies, metadata schemas 
--- can be a simple list of terms (properties, classes of resources), or a more 
complex ontology allowing inference and logics between terms. We know that a 
unique absolute ontology is utopia. Each SAs encode contextual knowledge, which 
need to be understandable, shareable and reusable by humans and machines.

Without fully implementing yet the standard semantic web toolkit, the 
astronomy communities are implementing open linked-data frameworks. They all
define resource classifications, propose community registries that provide 
links between the resources, and associate these with rich metadata. Upgrading 
these open science frameworks with semantic web technologies is then a matter 
of defining and mapping concepts without breaking the current solutions.

\subsection{Semantics Artefacts for FAIR Astronomy}

In Europe, the development of open data frameworks has been supported by the 
European Commission framework programs for many years. A significant number of 
astronomy projects have been funded throughout the last two decades. Table 
\ref{tab:grants} presents an already long (though non-exhaustive) list of such 
grants, ranked by starting years. The ESCAPE project, which ended in 2023, has 
initiated the merging of astronomy and particle physics infrastructures and 
standards.

\begin{table*}
{\small\begin{tabularx}{\linewidth}{lXllll}
\textbf{Acronym}
	&\textbf{Name}
	&\textbf{Topics}
	&\textbf{Identifier}
	&\textbf{Programme}
	&\textbf{Time range}\\\hline\hline
AVO
    &Astrophysical virtual observatory
    &A
    &\href{https://cordis.europa.eu/project/id/HPRI-CT-2001-50030}{HPRI-CT-2001-50030}
    &FP5
    &2001-2004\\\hline
EUROPLANET
    &European Planetology Network
    &PS
    &\href{https://cordis.europa.eu/project/id/1637}{1637}
    &FP6
    &2005-2008\\\hline
EUROVO-DCA
    &The European virtual observatory data centre alliance
    &A
    &\href{https://cordis.europa.eu/project/id/31675}{31675}
    &FP6
    &2006-2008\\\hline
EuroVO-AIDA
    &Euro-VO Astronomical Infrastructure for Data Access
    &A
    &\href{https://cordis.europa.eu/project/id/212104}{212104}
    &FP7
    &2008-2010\\\hline
EUROPLANET-RI
    &European Planetology Network Research Infrastructure
    &H,PS
    &\href{https://cordis.europa.eu/project/id/228319}{228319}
    &FP7
    &2009-2012\\\hline
HELIO
    &The Heliophysical Integrated Observatory
    &H
    &\href{https://cordis.europa.eu/project/id/238969}{238969}
    &FP7
    &2009-2012\\\hline
VAMDC
    &Virtual Atomic and Molecular Data Center
    &A,H,PS
    &\href{https://cordis.europa.eu/project/id/239108}{239108}
    &FP7
    &2009-2012\\\hline
CASSIS
    &Coordination Action for the integration of Solar System Infrastructures and
    Science
    &H,PS
    &\href{https://cordis.europa.eu/project/id/261618}{261618}
    &FP7
    &2010-2013\\\hline
EuroVO-ICE
    &Euro-VO International Cooperation Empowerment
    &A
    &\href{https://cordis.europa.eu/project/id/261541}{261541}
    &FP7
    &2010-2012\\\hline
IMPEx
    &Integrated Medium for Planetary Exploration
    &H
    &\href{https://cordis.europa.eu/project/id/262863}{262863}
    &FP7
    &2011-2015\\\hline
ESPAS
    &Near-Earth Space Data Infrastructure for e-Science
    &
    &\href{https://cordis.europa.eu/project/id/283676}{283676}
    &FP7
    &2011-2015\\\hline
ASTERICS
    &Astronomy ESFRI and Research Infrastructure Cluster
    &A
    &\hdoi{10.3030/653477}
    &H2020
    &2015-2019\\\hline
EPN2020-RI
    &Europlanet 2020 Research Infrastructure
    &H,PS
    &\hdoi{10.3030/654208}
    &H2020
    &2015-2019\\\hline
ESCAPE
    &European Science Cluster of Astronomy \& Particle physics ESFRI research 
    infrastructures
    &A,H,PP
    &\hdoi{10.3030/824064}
    &H2020 (EOSC)
    &2019-2023\\\hline
SOLARNET
    &Integrating High Resolution Solar Physics
    &H
    &\hdoi{10.3030/824135}
    &H2020
    &2019-2023\\\hline
EPN-2024-RI
    &Europlanet 2024 Research Infrastructure
    &H,PS
    &\hdoi{10.3030/871149}
    &H2020
    &2020-2024\\\hline
PITHIA-NRF
    &Plasmasphere Ionosphere Thermosphere Integrated Research Environment and 
    Access services: a Network of Research Facilities
    &H
    &\hdoi{10.3030/101007599}
    &H2020
    &2021-2025\\\hline
FAIR-IMPACT
    &Implementing FAIR data and services in the European Open Science Cloud
    &A,H,M,PS
    &\hdoi{10.3030/101057344}
    &HORIZON (EOSC)
    &2022-2025\\\hline
OSTRAILS
    &Open Science Plan-Track-Assess Pathways
    &A,H,M,PS,PP
    &\hdoi{10.3030/101130187}
    &HORIZON (EOSC)
    &2024-2017\\\hline
\end{tabularx}}
\caption{European commission grants including astronomy data management
developments and research infrastructures. The topics are listed with the 
following code: A - Astronomy; H - Heliophysics; M - Multi-disciplinary (outside 
astronomy); PP - Particle Physics; PS - Planetary Sciences. The identifiers 
before the H2020 programme are just links to the CORDIS web page for the
project.} \label{tab:grants}
\end{table*}

The scope of this study covers the astronomy community along with its three 
sub-domains (celestial astronomy, planetary sciences 
and heliophysics). Recent publications \citep[e.g.,][]{2022cosp...44.3491H,
2024arXiv241207973B,2024arXiv240103769C,2024ASPC..535..295L,2024ASPC..535..315S,
Masson:2024fc} show that each of these communities are claiming that they are 
producing a FAIR \citep[Findable, Accessible, Interoperable and 
Reusable,][]{Wilkinson:2016dn} ecosystem. The evaluation of the FAIRness of 
astronomy digital resources is discussed and studied by many groups, 
and is out of the scope of this paper. However, one of the FAIR 
criteria\protect\footnote{I2: ``(meta)data use vocabularies that follow FAIR 
principles''} explicitly requires that community vocabularies also complies 
with the FAIR principles. Moreover, the fact that each sub-community is still 
siloed is a strong hint that things can be improved. 

In 2018, the IVOA Solar System interest group started to bring 
heliophysics and planetary sciences needs in the IVOA ecosystem. The uptake of 
IVOA protocols in heliophysics and planetary sciences is growing, but at lot 
remains to be done, in particular on the semantic interoperability side, so that 
it is more widely adopted. This means that we need to harmonise our SAs, so that 
communities can exchange data and transform these data into knowledge --- 
semantically described, interoperable, actionable, and open. Moreover, the 
adoption of common SAs across communities will improve the FAIRness of 
data products, since the metadata would be described with the same terms, or 
with terms that are related one to the other. The increasing use of SAs in 
astronomy presents challenges including their description, selection, evaluation, 
trustworthiness and mapping. The landscape remains fragmented, with SAs scattered 
across various registries in diverse formats and structures --- not yet fully 
developed or encoded with rich semantic web standards based on RDF like OWL 
(Ontology for the Web Language, \citet{owl-ref}) or SKOS (Simple 
Knowledge Organisation System, \citet{skos-ref}) --- and often with overlapping 
scopes. There is thus need for a common place to catalogue, manage, assess, 
select and align FAIR SAs from several communities \citep{Gray:2015me,
Ritschel:2016bk,Rovetto:2023ao}. 

\section{Current semantic artefact management in astronomy}

The celestial astronomy community data alliance is the IVOA, which 
maintains an operational interoperability framework used by data repositories 
and science application platforms throughout the world. In this community, SAs 
are composed of vocabularies and schemas. The Semantics WG of the IVOA is 
managing the vocabularies used in the IVOA standards, 
following the \emph{Vocabularies in the VO} specification 
\citep{2023ivoa.spec.0206D}. The vocabularies are listed in a dedicated 
web page\footnote{\url{https://ivoa.net/rdf/}} where they are accessible using 
IVOA as well as with semantic web standard formats. 
Updates of the vocabularies are managed through a Vocabulary 
Enhancement Proposal (VEP) process \citep[for details, see][]{2023ivoa.spec.0206D}. 
The maintenance of the vocabularies is managed in a dedicated Github 
repository\footnote{\url{https://github.com/ivoa-std/Vocabularies}}. 
The Data Model WG of the IVOA is managing the schemas, following the \emph{Virtual 
Observatory Data Model Language} specification \citep{2018ivoa.spec.0910L}, but 
they are not available in semantic web compliant formats. The role on SAs in 
the IVOA is to ensure the interoperability, with controlled vocabularies to be 
used in protocol and metadata fields. In the recent years, most of the 
controlled valued lists, which were previously maintained in schemas by the Data 
Model WG, have been transferred to the Semantics WG. In this way, the IVOA SAs 
have already been prepared for a wider use.

In the planetary science community, two main frameworks co-exist with different 
scopes. The IPDA maintains an 
advanced data archiving metadata schema for planetary exploration datasets, the 
Planetary Data System version 4 information model\footnote{
\url{https://pds.nasa.gov/datastandards/}} (PDS4-IM) 
\citep{Hughes:2017bt}. The foundations of the PDS4-IM rely 
on the OAIS Standard (Open Archive Information System, ISO 14721:2012). It is 
developed using semantic web related tools, such as Prot\'eg\'e \citep{protege}, 
but is not implemented using RDF. The PDS4-IM is maintained and curated 
by a Change Control Board\protect\footnote{\url{https://pds-engineering.jpl.nasa.gov/pds4/ccb/}}, 
which is supported from a Data Design WG\protect\footnote{\url{https://pds-engineering.jpl.nasa.gov/pds4/ddwg/}}
representing the various PDS4 disciplinary nodes.
The Planetary Domain WG of the OGC (Open Geospatial Consortium) proposes a set 
of standards and tools used for the study of planetary surfaces. Although
the OGC already implements SAs in the form of RDF SAs (see, e.g., the 
ESIP ---Earth Science Information Partners--- Community Ontology 
Repository\protect\footnote{\url{http://cor.esipfed.org}}),
the Planetary Domain WG is not providing its SAs in RDF yet.

The heliophysics data ecosystem is managed by the IHDEA. This community is 
proposing a set of tools and standards for finding and accessing datasets for the 
heliophysics community. Semantic artefacts in this community were historically of 
three kinds: the SPASE Ontology (available as an XML schema), which includes lists 
of terms, properties and classes for defining various objects and resources 
(Persons, Observatories, Instruments, Datasets, Repositories, etc); the VSO (Virtual 
Solar Observatory) metadata; and the SOLARNET set of keywords (dedicated to 
Solar observations). Each of these three metadata standard are community 
effort to improve the findability and accessibility of their research products and 
their distributions. However, they are still currently not fully connected, and 
none of the them share an RDF implementation.

The IVOA, IPDA and IHDEA alliances are all worldwide 
collaborative groups, consensus 
and bottom-up driven, and based on best effort contributions. Interdisciplinary 
links between these communities have been mostly developed thanks to the 
Europlanet/VESPA\footnote{\url{http://www.europlanet-vespa.eu}} 
infrastructure \citep{Erard:2018kr}. They focus on discoverability and 
implementation of plugins to extend the capabilities of existing tools. The 
semantic interoperability across the sub-communities approach started only 
recently, with the ongoing development of three common SAs: a vocabulary for 
``observation facilities'' \citep{cecconi:2023of} and a proposal to 
the IHDEA to adopt two IVOA SAs for managing reference frames \citep{weigel:2025a} 
and units \citep{weigel:2025b}. 

In this paper, we present the results of a prototyping exercise. Thanks to the 
FAIR-IMPACT project, we have implemented an instance of the OntoPortal 
technology \citep{Jonquet:2023gb} to deploy a prototype SA catalogue for 
astronomy. Hereafter, we present current practices in astronomy, and review 
candidate SAs for such a cataloguing exercise. We report the use cases that have 
been driving this development. We present the first results, discuss the impact
and outcomes, and layout a roadmap for the deployment of a real production 
service for the astronomy community based on the OntoPortal technology, that 
will be called OntoPortal-Astro. Throughout the paper, the \protect\emph{astronomy 
OntoPortal instance} and the \protect\emph{OntoPortal-Astro} terms 
refer respectively to the prototype instance development within the FAIR-IMPACT 
project, and the upcoming official server.

Among the vocabularies that we included in the astronomy OntoPortal instance, we 
can list the following items: the Unified Astronomy 
Thesaurus\footnote{\url{https://astrothesaurus.org/}} \citep{Frey:2018ac} (UAT); 
the IVOA vocabularies (including properties and terms for the IVOA protocols 
interoperability); the EPNcore (Europlanet-VESPA core metdata) dictionary 
\citep{2022ivoa.spec.0822E}; the IHDEA community metadata including the SPASE 
schema\footnote{\url{https://spase-group.org/data/}}, the Virtual Solar 
Observatory data 
model\footnote{\url{http://docs.virtualsolar.org/wiki/DataModel18}}; the IUGONET 
schema\footnote{\url{http://www.iugonet.org/product/metadata.jsp}} 
(Inter-university Upper atmosphere Global Observation NETwork); the SOLARNET 
metadata \citep{solarnet_zenodo}; the CEF (Cluster Exchange Format) 
metadata \citep{allen:2009sh}; the IPDA information 
model\footnote{\url{https://pds.nasa.gov/datastandards/documents/im/}}. Other 
SAs will be included, such as: the VAMDC (Virtual Atomic and Molecular Data 
Centre) metadata schema\footnote{\url{http://dictionary.vamdc.eu/}}, the OGC 
planetary frame registry\footnote{\url{http://voparis-vespa-crs.obspm.fr}}. The 
GEMET (GEneral Multilingual Environmental Thesaurus) 
thesaurus\footnote{\url{https://www.eionet.europa.eu/gemet/}} could also be 
included, however we might federate with to EarthPortal (Earth sciences 
community OntoPortal instance) for this SA. Regarding particle physics, the CERN 
Open Data\footnote{\url{https://github.com/cernopendata}} terms and schemas will 
also be included.

The two main challenges of this project are: (i) to
identify the SAs, which are currently scattered in many places, and (ii) to 
produce SAs in a standard semantic web form (typically OWL, or SKOS). Besides being a 
requirement for being ingested into the astronomy OntoPortal instance, it is 
also a logical evolution for our artefacts. Indeed, most of the current SAs 
(except those from the celestial astronomy community) are in diverse forms, from 
lists of terms embedded in XML schemas, to unformatted lists of metadata in 
specification documents (PDF files, HTML pages, etc.) This ``semantic lifting'' 
shall be done by the semantics-related working groups or authorities of 
the relevant 
communities (e.g., the IVOA Semantics WG or the IHDEA dedicated teams), with the 
support of expert in semantics and partners from the OntoPortal Alliance 
(mostly from the AgroPortal team, the Agri-food community OntoPortal instance 
developed by INRAE) as well as from the OntoPortal-Astro team (see 
Section \ref{sec:disc}).

\section{Use Cases}
A series of use cases have been identified to trigger the developments of this 
SA catalogue dedicated to the astronomy communities.

\subsection{Exoplanets}
Exoplanetary objects (defined as planets orbiting around other stars than 
the Sun) and their environments are studied with increasing resolution 
and capabilities. Comparisons with solar system object observations become very 
promising. Solar system observations should then be available and semantically 
compliant with the IVOA astronomy ecosystem. Currently, the IVOA object type 
vocabulary (\texttt{objtyp}, see Table \ref{table:ns} for all namespace 
definitions) lists the term \objtyp{planet} defined as an ``Extra-solar 
Planet" (not including the solar system objects). On the other hand, the EPNcore 
dictionary includes a list of target classes (\texttt{tarcls}), where the 
terms \tarcls{planet} and \tarcls{exoplanet} respectively refer to ``solar 
system planets'' and ``exoplanets''. Similarly the UAT (\texttt{uat}) proposes 
two concepts: \uat{1260} and \uat{498} for ``Solar system planet'' 
and ``Exoplanets'' respectively. The IVOA UAT vocabulary (\texttt{ivoauat}) 
mirrors the UAT terms that thus also proposes the terms 
\ivoauat{solar-system-planet} and \ivoauat{exoplanets}. This requires careful 
management of the metadata across domains, with adequate mappings.  

Currently, the \texttt{uat} and \texttt{ivoauat} thesauri are published as SKOS 
vocabularies. The \texttt{objtyp} and \texttt{tarcls} vocabularies are draft SAs 
prepared respectively by the IVOA Semantics WG, and by 
the Solar System interest group of the IVOA with the Europlanet team, and are 
being developed as OWL ontologies, using either Prot\'eg\'e, or the IVOA
vocabulary framework. Mapping these SAs is not a difficult task, but 
collecting them into a single OntoPortal instance makes it easier.

The current IVOA use of vocabularies is strongly implicit: the IVOA 
specifications list which SA should be used in controlled-valued fields of 
protocols and interfaces. The IVOA client software are built with this implicit 
knowledge, which is enabling a technical interoperability. The semantic mapping 
between terms, if required for a use case, have thus to be implemented in 
clients, upstream from the IVOA interfaces and protocol queries. The mappings 
are currently stored in the IVOA vocabularies: e.g., in the IVOA UAT, each term 
is semantically mapped to its twin concept in the UAT thesaurus, using 
a \skos{exactMatch} property.

\subsection{Space Weather}
A widely accepted definition of Space Weather is: ``\emph{Space Weather is the 
physical and phenomenological state of natural space environments. The 
associated discipline aims, through observation, monitoring, analysis and 
modelling, at understanding and predicting the state of the Sun, the 
interplanetary and planetary environments, and the solar and non-solar driven 
perturbations that affect them; and also at forecasting and nowcasting the 
possible impacts on biological and technological systems}'' 
\citep{Lilensten:2009}. Space weather is thus about heliophysics and 
its impact to the Earth environment. Space Weather services are being developed 
by ``classical'' weather organisations, such as the Met Office (UK), using 
mainly Earth observation software infrastructure (OGC), whereas the original 
data are produced by the heliophysics and astronomy communities (IHDEA and IVOA 
ecosystems). Collating and mapping SAs between OGC and the astronomy data 
ecosystems is required to efficiently exchange information, including mappings 
and coordination on the development of common ontologies. This implies mapping 
terms about reference frames, units of measure, physical quantities, sensor 
descriptions, etc.

\subsection{Observation context}
Modern instruments are getting very versatile with many modes of operations, 
thanks to the increasing digital processing capabilities, closer to the data 
acquisition devices, and implying automated data processing. Provenance metadata 
are increasingly implemented to trace these processing steps, requiring 
ontologies to efficiently model observational and instrumental contexts. 
Astronomy, heliophysics or particle physics ontologies need to be aligned and 
used in a coordinated manner to allow reuse in interdisciplinary studies. 

The W3C has adopted a data model (Prov-DM) \citep{w3c-prov-dm} and an ontology 
(Prov-O) \citep{w3c-prov-o} for managing provenance metadata. The IVOA inspired 
from those recommendations with the VO-PROV data model \citep{vo-prov}, 
extending the W3C Prov-DM. The IPDA and IHDEA have not yet standardised such 
metadata. Mapping VO-PROV to Prov-DM (and thus Prov-O) is 
straightforward, but should be implemented. We should also explore how the IPDA 
and IHDEA teams are individually managing their provenance metadata when 
applicable. Note that for the general domain-agnostic semantic web 
vocabularies, the OntoPortal FAIR-IMPACT partners are currently working on a 
specific portal (LovPortal) to host the content of the Linked Open Vocabulary
platform.

\subsection{Harmonisation of nomenclature}
Some concepts are used by all sub-communities of astronomy, such as the observed
physical quantities, the observation facilities, the instrument types, or the 
coordinate reference frames. Harmonisation of these generic terms and concepts 
is needed for semantic interoperability. Such developments already started for 
observed parameters \citep{cecconi:2014ucd}, observation facilities 
\citep{cecconi:2023of}, as well as coordinate reference frames in an IHDEA
working group. For observation facilities and reference frames, the current
plan is to let the IVOA Semantics WG manage the merged vocabularies.  
The astronomy community OntoPortal instance will 
provide a platform to follow up and further develop these projects. 
Within OntoPortal every SA is automatically mapped one-another (at the lexical 
level) and term reuses are explicitly exposed and detected by the tool.

\subsection{Scientific Knowledge Graphs} 
Connecting research outputs to persons, institutions and funders is a key task
for enabling the FAIR principles, discovering research products or measuring 
the scientific impact of the projects and infrastructures. Several tools are 
implementing so called SKG, which can mean either Scholarly Knowledge Graph
(such as with the OpenAIRE Graph 
infrastructure\footnote{\url{https://graph.openaire.eu}}), or Scientific 
Knowledge Graph (such as with the SciLake 
project\footnote{\url{https://scilake.eu}}). Moreover, the astronomy 
community registries, such as the IVOA registry or the SPASE registry, can also
be named SKGs as they relate all the elements of the research ecosystem (data
products, repositories, scholarly articles, institutions, persons, etc.) 

Building efficient and FAIR knowledge graphs requires to use well-defined and 
cross-domain ontologies, so 
that term URIs can be used as search keys in queries (see, e.g., the 
``exoplanet'' use case). OntoPortal-Astro will implements a 
very rich metadata model to describe SAs --- based on the MOD metadata 
vocabulary\footnote{\url{https://github.com/FAIR-IMPACT/MOD}}
\citep{Dutta:2017} --- which will support the formal and semantic description of 
astronomy SAs in SKGs. Plus, it will embed a 
FAIRness assessment tool based on the O'FAIRe methodology \citep[Ontology 
FAIRness Evaluation,][]{Amdouni:2022} to help in the selection of SAs. 
OntoPortal-Astro will thus facilitate the reuse of astronomy SAs making 
them more findable.

\subsection{Models and observations} 
Astronomy is an observation-based science, since the environmental conditions
of the observed objects and phenomena can't be controlled. It is thus necessary 
to try and model the possible environmental conditions leading to the 
observations. Within the IVOA, the celestial astronomy community has developed 
the Simulation Data Model \citep{2012:sim-dm}, which can describe the 
astrophysical simulation parameters. However, due to its complexity, it has 
not been adopted by many teams. In the heliophysics community, the need to 
coupling measurements to simulations has been identified as a key issue 
\citep{Masson:2024fc}, either to discover and reuse observational data as input 
for environment models, or to compare observations to output parameters of those 
models. Hence, the observed and modelled physical parameters must be described 
the same way. The same reasoning applies to planetary environment, such as the 
global atmospheric models of planets, or the planetary magnetosphere models 
\citep[see, e.g.,][]{Erard:2018kr,Achilleos:2019if,Azria:2021jk,Trompet:2022ll}. 

Describing observable properties (or 
scientific variables) within ontologies or other types of SAs, is
a growing practice. It is present in multiple 
OntoPortal community instances such as Earth sciences (EarthPortal),
Agri-food (AgroPortal) or Ecology (EcoPortal), using the Extensible 
Observation Ontology \citep[OBOE,][]{Madin-oboe:2007} or the InteroperAble 
Descriptions of Observable Property Terminology\footnote{\url{https://i-adopt.github.io}} 
(I-ADOPT)). For instance, in 
EcoPortal, the ENVTHES (Environmental Thesaurus) already use the I-ADOPT 
representation to encode their scientific variable within these SAs
\footnote{See, e.g., \url{http://ecoportal.lifewatch.eu/ontologies/ENVTHES/30209}}.

\section{First Results}
Based on the use-cases presented in the previous section, work has been 
initiated mostly in the frame of the FAIR-IMPACT project. 

\subsection{Review of pre-exisiting Semantic Artefact Catalogues}

A recent landscape analysis of catalogues containing SAs has identified 
three among the astronomy community: the IVOA vocabularies, the NASA/PDS data 
standards and the ESPAS Vocabulary broker \citep{jonquet_2024_12799862}. SA 
catalogues have been classified according to their typology and the level of 
service offered. A \emph{listing} corresponds to a simple 
online listing of SAs, while a \emph{library} represents structured 
online listing and the a \emph{repository} is considered as 
an advanced web application \citep{jonquet_2024_12799796}.
The IVOA vocabulary\footnote{\url{https://ivoa.net/rdf/}} web page is considered 
as a SA library, which gathers the vocabularies developed and maintained by the 
IVOA Semantics WG. The NASA/PDS data 
standard\footnote{\url{https://pds.nasa.gov/datastandards}}
web page is a SA listing, which provides all version of the NASA/PDS information 
model, and the associated vocabularies. The ESPAS Vocabulary Broker, which is
consider as retired \citep{jonquet_2024_12799796}, and is not accessible anymore 
at the time of writing, is a SA repository. It was providing access to the terms 
from the SPASE schema, the ESPAS project ontology, the GEMET ontology and the 
UAT thesaurus. Moreover, the ESPAS vocabulary used to offer mapping between 
terms of ontologies among the repository. Among those, solely the IVOA 
vocabulary library partly meets the SA catalogues FAIRness criteria proposed in 
\citet{jonquet_2024_12799796}. Hence, no pre-existing solution satisfies the 
needs for a common astronomy SA catalogue. There is thus a need of harmonising 
and synchronising the different initiatives to avoid technical and semantic 
silos.

\subsection{OntoPortal-Astro, an astronomy community OntoPortal}
In the FAIR-IMPACT projet, an astronomy SA catalogue use-case has been 
developed \citep{cecconi:2024lm}. An OntoPortal instance for astronomy 
has been installed:\\
\url{https://ontoportal-astro.eu}\footnote{At the time of writing,
this server presents the astronomy OntoPortal prototype developed during the 
FAIR-IMPACT project, but will be replaced by the OntoPortal-Astro within 
a few months.}\\
The goal of the prototype was two-fold: testing the technology to expose, and 
assess SAs in astronomy. We selected about 40 vocabularies and ontologies that 
we would use as tests of the usage of the framework, see Table 
\ref{tab:ontoportal-examples}. In addition to all the IVOA Vocabularies, we also 
drafted OWL ontology versions of several community SAs (three schemas and one 
vocabulary): (i) the NASA/PDS4 information model, (ii) the IHDEA/SPASE data 
model, (iii) the VAMDC schema, and (iv) the CERN Open Data terms. The 
development of an ontology version of a schema is more than a format conversion, 
since it implies a full modelling of the concepts, classes and properties, as 
well as minting URIs for the terms. Hence, the development of these SAs have 
been used to exercise the astronomy team members. We interacted with the schema 
maintainers to share our explorations. The NASA/PDS4, IHDEA/SPASE and VAMDC 
teams expressed a strong interest in our approach. In the future, we envision 
the sharing in OntoPortal-Astro as part of the life cycle of each of these SAs 
as it is advocate by classic vocabulary development methodologies such as Linked 
Open Terms \citep[LOT,][]{POVEDAVILLALON2022104755} that has 
also include FAIR assessment steps to ensure the developed artefacts are FAIR 
by design and appropriately shared.

\begin{table*}
\small\begin{tabularx}{\textwidth}{llllX}
\textbf{Name}
	&\textbf{Metrics}
	&\textbf{Topics}
	&\textbf{Status}
	&\textbf{Notes}\\\hline\hline
Astronomical Subject Keywords
	&Concepts: 394
	&A,H,PS
	&Test
	&Old ontology developed by the AAS (American Astronomical Society), still
	used by some journals.\\ \hline
CDPP TREPS tool Reference frames
	&Classes: 78
	&H,PS
	&Develop
	&Intermediate ontology extracted from \href{http://treps.irap.omp.eu}{TREPS}
	tool source code. To be merged into IVOA RefFrame\\\hline
EPNcore Dataproduct Types 
	&Classes: 14
	&H,PS
	&Develop
	&EPNcore v2.0 Dataproduct Type vocabulary.\\\hline
EPNcore Small Body Dynamical Classes
	&Classes: 19
	&PS
	&Develop
	&EPNcore v2.0 Small Body Dynamical Class vocabulary.\\\hline
EPNcore Spatial Frame Types
	&Classes: 7
	&H,PS
	&Develop
	&EPNcore v2.0 Spatial Frame Type vocabulary.\\\hline
EPNcore Target Classes
	&Classes: 13
	&H,PS
	&Develop
	&EPNcore v2.0 Target Class vocabulary.\\\hline
GEMET Thesaurus
	&Concepts: 5739
	&H,PS
	&Test
	&Test import of the GEMET Thesaurus. To be coordinated with 
	EarthPortal.\\\hline
IVOA Content types of VO resources
	&Classes: 22
	&M
	&Ready
	&Generic types of resources in the Virtual Observatory.\\\hline
IVOA Datalink core
	&Properties: 22
	&M
	&Ready
	&Properties linking described object to other resources, in the DataLink 
	IVOA 
	standard.\\\hline
IVOA Object Types
	&Classes: 151
	&A
	&Develop
	&Astronomical object types used in the IVOA, still in development.\\\hline
IVOA Messengers
	&Classes: 9
	&A,H,PS
	&Ready
	&Messenger (type of particule) providing the information observed in 
	astronomy.\\\hline
IVOA Product Types
	&Concepts: 20
	&A
	&Ready
	&Data Product type vocabulary for the IVOA.\\\hline
IVOA Reference Positions
	&Classes: 6
	&A
	&Ready
	&Reference positions used in the IVOA.\\\hline
IVOA Relation Types
	&Properties: 14
	&M
	&Ready
	&Used for relations between resources in the IVOA registry\\\hline
IVOA Roles of dates
	&Properties: 14
	&M
	&Ready
	&Roles of dates in the IVOA Registry\\\hline
IVOA Semantics
	&Properties: 4
	&M
	&Ready
	&This vocabulary defines several properties used in the management of IVOA
	vocabularies.\\\hline
IVOA Time Scales
	&Classes: 9
	&A,H,PS
	&Ready
	&Time scales used in the IVOA.\\\hline
IVOA Unified Astronomy Thesaurus
	&Concepts: 2373
	&A,H,PS
	&Ready
	&IVOA Rendering of Unified Astronomy Thesaurus (UAT)\\\hline
IVOA Unified Content Descriptors (UCD) v1.5
	&\begin{minipage}[t]{2.5cm}
	Classes: 6\\
	Individuals: 554
	\end{minipage}
	&M,A,H,PS
	&Develop
	&Rendering of IVOA UCD vocabulary as an ontology.\\\hline
NASA NAIF Spacecraft Codes
	&Classes: 150
	&A,H,PS
	&Test
	&Test rendering of the NASA NAIF spacecraft code list as on 
	ontology.\\\hline
NASA PDS4 Information Model
	&\begin{minipage}[t]{2.5cm}
	Classes: 221\\
	Properties: 114
	\end{minipage}
	&PS
	&Test
	&Ontology version of the PDS4 Information Model\\\hline
NASA PDS4 Information Model (Full dump)
	&Classes: 1038
	&PS
	&Test
	&Test rendering of the NASA PDS4 Information Model as an ontology (full dump 
	provided by the NASA team).\\\hline
NASA PDS4 (from ESIP)
	&Classes: 1149
	&PS
	&Test
	&PDS4 information mode rendered as on ontology by the ESIP group.\\\hline
PDS Instrument Types
	&\begin{minipage}[t]{2.5cm}
	Classes: 186\\
	Individuals: 8
	\end{minipage}
	&PS
	&Test
	&Test ontology of instrument types, from various IPDA sources.\\\hline
SPASE Reference frames
	&Classes: 47
	&H
	&Develop
	&Intermediate ontology extracted from the SPASE XML Schema. To be 
	merged into IVOA RefFrame.\\\hline
SPASE OWL Test Implementation
	&\begin{minipage}[t]{2.5cm}
	Classes: 129\\
	Individuals: 642\\
	Properties: 273
	\end{minipage}
	&M,H
	&Develop
	&Test rendering of the SPASE XML schema as an ontology.\\\hline
Unified Astronomy Thesaurus (UAT)
	&Concepts: 2373
	&A,H,PS
	&Ready
	&Used for keywords in astronomy papers\\\hline
VAMDC Vocabulary
	&\begin{minipage}[t]{2.5cm}
	Classes: 603\\
	Properties: 554
	\end{minipage}
	&M,A,H,PS
	&Develop
	&Test rendering of the VAMDC schema as an ontology.\\\hline
\end{tabularx}
\caption{Subset of the list of semantic artefacts implemented in the astronomy 
OntoPortal instance prototyped during FAIR-IMPACT. The topics column letter 
codes are: A -- Astronomy; H -- Heliophysics; M -- Metadata; PS -- Planetary 
Sciences. The status can be: \emph{Test} (for testing purposes); \emph{Develop} 
(currently being developed); \emph{Ready} (already in use).} 
\label{tab:ontoportal-examples}
\end{table*}

\subsection{Improvement of IVOA vocabularies}\label{sec:improve}
Importing the IVOA vocabularies into our astronomy OntoPortal instance revealed 
minor design issues in the IVOA vocabularies management and their associated RDF 
artefacts, as defined in \citet{2023ivoa.spec.0206D}. We fixed the SKOS 
vocabularies, which were missing top level concepts. This prevented tools to 
correctly build the concept hierarchies out of the ontology. We also fixed the 
way the URI were referenced internally in the ontologies. Finally, we started to 
add more metadata to our vocabularies, such as, preferred namespace prefix and 
URIs, using the multiple properties harvested and promoted by MOD. We also 
initiated a mapping exercise between some of our properties with more commonly 
used ones. For instance, we have two properties used to handle deprecated terms 
and redirect to a replacement term: \ivoasem{deprecated} and 
\ivoasem{useInstead}, as part of our IVOA Semantics vocabulary (IVOASEM). 

\subsection{Strong heliophysics support}
Thanks to the Transform to Open Science (TOPS) project, the NASA contributors
to IHDEA started several projects aiming at improving semantic interoperability
of the heliophysics community. A first example is the conversion of the SPASE 
XML Schema into an OWL 
ontology\footnote{\url{https://github.com/polyneme/topst-spase-rdf-tools}}.
A second ongoing work is the mapping of the SPASE XML Schema to 
schema.org\footnote{\url{https://github.com/clnsmth/soso}}, in
coordination with \emph{Science on Schema.org} 
\citep{adam_shepherd_2024_7884538,ringuette_2025_15594513}. When 
implemented and operational, these mappings will require an easy access 
to SAs, based search interfaces on ontology terms, so that term URIs are 
seamlessly inserted in the metadata. OntoPortal-Astro will provide a search 
interfaces for each SA, including a programmatic access.

\section{Challenges and impact}
As presented in the previous sections, the astronomy community is composed of 
several sub-communities (celestial astronomy, planetary sciences and 
heliophysics). Each community maintains and structures their SAs in 
heterogeneous forms. Since their drivers are also different, they have various 
maturity level concerning the management of SAs. They also store their 
ontologies in various repositories, using different technologies. However, the 
main goal of all the SAs is to enhance the FAIRness of research products (like,
e.g., published datasets).

\begin{figure*}
\centering\includegraphics[width=0.8\linewidth]{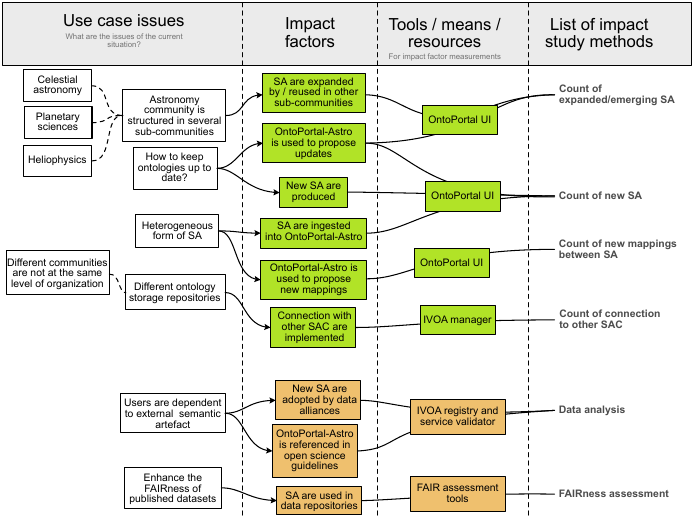}
\caption{Impact factors and metrics for assessing the impact of the OntoPortal 
technology on the astronomy community. Figure adapted from 
\citet{fair-impact-m4.5}.} \label{fig:impact}
\end{figure*}

The OntoPortal technology has been identified to improve this 
situation, by bringing all SAs into a common catalogue with advanced 
capabilities. Nine impact factors have been identified by the FAIR-IMPACT 
team, which are directly linked to the adoption of the OntoPortal 
technology, and seven metrics are proposed to monitor the impact. The impact 
factors and metrics are listed in Figure \ref{fig:impact}.

Most of the proposed metrics are based on monitoring the 
evolution of SAs using OntoPortal-Astro: (a) number of expanded or emerging SA; 
(b) number of new SAs; (c) number of new mappings between SAs; and (d) the 
number of connection with other SA catalogues. The adoption of SAs by data 
alliances could also be assessed through OntoPortal-Astro, if the metadata
include the list of projects the SAs are used in. The adoption of SAs in
data repositories requires to browse the community repositories to assess
the presence in the repository record metadata of URIs emanating from SAs 
referenced in OntoPortal-Astro. For the last two assessment methods, the 
developments done within the SKG-IF (Scientific Knowledge Graph 
Interoperability Framework) working group of the Research Data Alliance (RDA)
will provide a way to access repositories and registries homogeneously, 
facilitating thus the monitoring of, e.g., \dct{subjects} URIs.

\section{Discussion}
\label{sec:disc}
The development of the astronomy OntoPortal prototype has been 
truly interdisciplinary. It was enabled by the FAIR-IMPACT project, which 
has been funded European Open Science Cloud (EOSC) infrastructure 
programme. One of the goals of FAIR-IMPACT was to develop a 
multi-disciplinary federation of SA catalogues. The project selected the 
OntoPortal technology. Although the astronomy communities are mature 
communities, with well-established data ecosystems and solid bottom-up 
metadata governance, their use of semantic web technologies is not so much 
developed, compared to the practices in place in other communities like 
environmental or life sciences. The collaboration between the astronomy 
community on the one hand, and INRAE and the OntoPortal alliance on the other 
hand, brought astronomy the missing pieces to fast-forward to the modern age of 
web-semantics, with state-of-art technologies. The collaborations 
initiated during this project will be continued, with upcoming activities on 
OntoPortal-Astro and other projects of the EOSC programme.

As shown in \citet{Molinaro:2020}, in the frame of the ESCAPE 
project, the IVOA Registry has been connected to B2Find (a service provided by 
EUDAT, in the frame of EOSC): \url{https://b2find.eudat.eu/organization/ivoa}. 
This provides a new venue\footnote{In the sense of the SKG-IF ``venue'' 
definition: \emph{A publishing ``gateway'' used by an Agent to make their 
Research products available to others.}} for finding IVOA services and 
collections. However, the current connection is not implementing any of the 
implicit IVOA semantics (like, e.g., the IVOA UAT keywords, which are 
available as pure text terms, and not URIs). The goal of the FAIR-IMPACT 
project was to expand FAIR solutions across EOSC, including semantic web 
technologies. Once OntoPortal-Astro is operational, further work with the 
IVOA and EUDAT/B2Find will have to be conducted to connect this new SA 
catalogue to the B2Find search portal.

The developments around the astronomy OntoPortal prototype have been regularly
discussed within the IVOA Semantics WG. While the WG members are fully aware 
and positive about the open 
science challenges and the need for improving the astronomy SAs, the common 
ground is that the IVOA vocabularies are tailored for enabling interoperability 
within the IVOA. Hence, any development on the IVOA vocabularies shall be made 
backward compatible with the current uses and practices and shall not impose 
extra burden on the maintainers, tool developers and users of the vocabularies. 

The IVOA vocabularies are setup so that they have as little external 
dependencies as possible. The main example is the UAT, which
is developed independently from the IVOA. The \emph{Unified Astronomy Thesaurus 
(IVOA rendering)} is a regularly updated copy of the UAT thesaurus. 
The IVOA version includes the relations to the original UAT terms, so that 
linked data applications can connect resources using the same terms. However, in 
the unfortunate event of discontinuation of the UAT reference thesaurus, the 
IVOA interfaces based on those terms will continue to work. The current design 
is thus trying to keep a strong infrastructure, as well as allowing linked and 
open data applications. For this, OntoPortal-Astro will 
serve as a long term solution 
in case a vocabulary cease to be supported (e.g., URIs are not resolvable 
anymore), as the technology include a URI resolution and negotiation service for 
each hosted SA.

Even with the impulse of OntoPortal-Astro, the semantic artefact 
management may not change in all communities immediately. For instance the IVOA 
has already a well established SA governance, and the improvement should only be 
limited to the improvement of the SA metadata. For planetary sciences and 
heliophysics, the first step will be to include OWL or SKOS versions of their 
SAs, and then include this scheme into their SA governance. Currently, the 
heliophysics community is studying the mapping of their registry metadata records
to schema.org, which requires to produce SAs and accessible concept URIs. In the 
planetary science community, the IPDA is currently not ready for such a shift, 
but the planetary surfaces OGC group should be able to produce SAs related to 
coordinate systems and projections.
 
Improving the FAIRness of the astronomy vocabulary is also one of the key 
outcome of this study. We used FAIR assessment tools dedicated to ontologies
and SAs, like 
FOOPS!\footnote{\url{https://foops.linkeddata.es/FAIR_validator.html}} 
\citep{foops2021}. Among all the assessment criteria, we can list the following
aspects. The term URIs should be persistent, but the tools only consider a 
limited set of http domains as persistent. The metadata should include a list 
of contributors, for the ontology and the individual terms. It should also 
include provenance metadata. A versioning schema should also be in place,
as discussed in \citet{jonquet:hal-04094847}. A 
preferred prefix should be advertised. Some criteria can easily be set up, and 
others (like the persistent identifier one) would require coordination with the 
FAIR assessment tool maintainers. On this aspect, the future OntoPortal-Astro 
will include O'FAIRe tool which provides a nice reporting on how to improve the 
FAIRness of each SAs.

This study is a first step towards improved management practices of SAs for 
astronomy. We focused here on vocabularies and ontologies, which
are already in use by the astronomy communities. A further step is to ease and 
extend their usage for describing and annotating research objects. The 
astronomy community pilot developed within the OSTrails project will make 
use of the FAIR SAs developed using OntoPortal for connecting 
resources through Science Knowledge Graph interfaces. Finally, another 
framework identified for annotated research products with description of 
measurements is I-ADOPT. It is proposing a simple template with design patterns 
and is making use of existing ontologies. We participated to the I-ADOPT 
Variable Modelling Challenge in 2024, with an example variable ``electron 
density in the solar wind''. From this simple test, the framework seems 
applicable, but, a lot concepts related to astronomy are missing.
We will explore more examples with astronomy use 
cases and identify which ontologies and concepts should be implemented 
to cover the needs. 

Keeping our astronomy ecosystem stable while opening up to linked data 
practices requires to build term mappings. Recently developed tools \citep[see, 
e.g., MSCR -- Metadata Schema and Crosswalk Registry, ][]{EUNIS2024:MSCR} are 
promising resources for storing mappings, including more complex conditional 
schema mappings than the current capabilities of OntoPortal, as those developed 
during the SPASE XML Schema to Schema.org exercise. The mappings will 
then have to be stored, either in independent mapping registries (such as the 
MSCR or the native OntoPortal mapping repository) or integrated into our 
ontologies. The use of artificial intelligence and large 
language models will also be 
tested for aligning vocabularies and building mappings. 

Since January 2025, the OSCARS cascading grant ``Ontology Portal for Astronomy 
Linked-data''\footnote{\url{https://oscars-project.eu/projects/opal-ontology-portal-astronomy-linked-data}} 
(OPAL) started. Its goal is to build a real production service, OntoPortal-Astro 
for the astronomy community building on the preliminary results from 
FAIR-IMPACT. Through this project, we plan to gather, update and produce SAs 
for all the astronomy sub-communities. We will also explore how to 
better expose our SAs so that they can be reused. The OntoPortal-Astro
team will include ontologists and knowledge designers, and will be supported by
an Advisory and User Group composed of experts from the various astronomy 
communities. OntoPortal-Astro will join 
the Ontoportal Alliance, and the portal will federate with other 
OntoPortal instances, such as EarthPortal, which is serving ontologies we can 
relate our terms to (for the space weather community or the planetary surfaces). 
We also consider developing domain ontologies, which will allow us to annotate 
our research products in a more accurate manner.

\section{Conclusion}

OntoPortal-Astro may become the reference Semantic Artefact Catalogue for 
astronomy, it shall represent a significant step toward strengthening the 
semantic infrastructure in astronomy by providing a centralised service for FAIR
astronomy SAs. By integrating and addressing the challenges from heliophysics to 
planetary sciences and across related astronomy disciplines, it will foster 
greater interoperability, data discoverability, and reuse. Through collaboration 
with the OntoPortal Alliance, this initiative will not only enhance the 
accessibility and management of SAs but also support interdisciplinary research 
across astronomy and beyond. Ultimately, OntoPortal-Astro will contribute to a 
more cohesive and FAIR-aligned semantic ecosystem, benefiting the broader 
scientific community.
 
\section*{Author Contributions}
Baptiste Cecconi (IVOA, IPDA, IHDEA and FAIR-IMPACT): 
\emph{Writing - Original Draft, Conceptualization,  Investigation, Methodology, 
Data Curation, Validation, Supervision}. 
Laura Debisschop  (FAIR-IMPACT): \emph{Writing - Review \& Editing, Validation, 
Investigation, Data Curation}. 
S\'ebastien Derri\`ere  (IVOA): \emph{Writing - Review \& Editing, 
Conceptualization, Validation}. 
Mireille Louys (IVOA): \emph{Writing - Review \& Editing, Conceptualization, 
Validation}.
Cl\'ement Jonquet (OntoPortal, FAIR-IMPACT): \emph{Writing - Review \& Editing, 
Project administration, Methodology, Software, Supervision}.
Carmen Corre (FAIR-IMPACT): \emph{Writing - Review \& Editing, Methodology, 
Project administration}.
Nina Grau (FAIR-IMPACT): \emph{Writing - Review \& Editing, Project 
administration, Methodology, Conceptualization, Data Curation}.

\section*{Acknowledgments}
This work has been supported by the FAIR-IMPACT project, which received funding  
from the European Commission's Horizon Europe Research and Innovation programme 
under grant agreement number 101057344. The authors acknowledge the Observatoire 
de Paris-PSL for hosting the OntoPortal prototype used for this work and the 
OntoPortal Alliance for their support and technology. The authors also thank 
Markus Demleitner (former chair of the IVOA Semantics WG), as well as the IVOA, 
IHDEA and IPDA groups for their helpful discussions
and feedback. The authors also acknowledge support from Sophie Aubin from INRAE.
 
\appendix
\setcounter{table}{0}
\section{Semantic artefact namespaces}
Table \ref{table:ns} lists the semantic artefact namespaces used throughout the 
paper. 
\begin{table*}
\small\begin{tabularx}{\textwidth}{Xlll}
SA Name&Namespace&URI&Topics\\\hline 
IVOA Object Types Vocabulary
	&\texttt{objtyp}
	&\url{\nsobjtyp}
	&A\\
EPNcore Target Classes 
	&\texttt{tarcls}
	&\url{\nstarcls}
	&P\\
Unified Astronomy Thesaurus
	&\texttt{uat}
	&\url{\nsuat}
	&A,P,H\\
IVOA UAT Vocabulary
	&\texttt{ivoauat}
	&\url{\nsivoauat}
	&A,P,H\\
Simple Knowledge Organisation Systems (SKOS)
	&\texttt{skos}
	&\url{\nsskos}
	&M\\
IVOA Semantics Vocabulary
	&\texttt{ivoasem}
	&\url{\nsivoasem}
	&M\\
Dublin Core Metadata Initiative (DCMI) Metadata Terms
	&\texttt{dct}
	&\url{\nsdct}
	&M
\end{tabularx}
\caption{List of namespaces used in this paper. The topics code letters are the 
same as in Table \ref{tab:ontoportal-examples}.}\label{table:ns}
\end{table*}

\bibliographystyle{elsarticle-harv}
\bibliography{semartastro}

\end{document}